# A Multilingual on-line Dictionary of Astronomical Concepts

M. Heydari-Malayeri
LERMA, Observatoire de Paris

**Abstract.** On the occasion of the International Year of Astronomy (IYA2009), we present a new *interactive* dictionary of astronomy and astrophysics, which contains about 7000 entries. This interdisciplinary and multicultural work is intended for professional and amateur astronomers, university students in astrophysics, as well as terminologists and linguists. A new approach is pursued in the formation of a scientific dictionary, which aims to display additional dimensions of astronomical concepts. Although Virtual Observatories recognize the necessity of efforts to define basic astronomical concepts and establish their reciprocal relations, so far they have mainly been confined to archiving observational data. The present dictionary could be an incipient contribution to cover and inter-relate the whole astronomical lexicon beyond subfields.

## 1. Prologue

On the occasion of the International Year of Astronomy (IYA2009), an almost complete version of the online *Etymological Dictionary of Astronomy and Astrophysics* is now available [1].

This newly installed *interactive version*, probably the largest compiled astronomical vocabulary, already contains about 7000 entries. In fact this work turns out to be the first etymological/linguistic/terminological dictionary ever created in astronomy/astrophysics. Since astronomy is tightly related to other branches of knowledge, the dictionary also includes a large number of terms in physics, mathematics, geology, meteorology, and even philosophy, among others.

It is organized as a Mysql/Php5 database to ease the search by key words. Moreover, the hypertext ability enables the reader to move on from a given concept to related ones. The pedagogical/educational aim is to provide broad information on the various aspects of the topic under consideration.

The rationale, principles, and criteria upon which this project lies are exposed in the *Introduction* of the online dictionary [1]. This interdisciplinary and multicultural work is intended for professional and amateur astronomers, university students in astrophysics, as well as terminologists and linguists, in particular those concerned with the Indo-European languages.

## 2. Entry structure, etymology, astronomy linguistics

In practice, the dictionary presents the definition of classical as well as advanced concepts of modern astronomy. The main entries, in English, are accompanied by their French and Persian equivalents in the present version. The Persian equivalents appear twice, in transliterated Latin characters and in proper Persian writing. The first paragraph of each entry block is devoted to the definition(s) of the entry. The first statement of the definition presents a concise, but explicit description of the concept. The following sentences elaborate on the information and lead the reader, through hyperlinks, to complementary explanations. The second paragraph describes the etymology of the English term. Indeed, the origin, history, and the way in which the term is composed provide the reader with an additional dimension of the concept. This section is in fact the interface between physical sciences and linguistics. The third paragraph presents the etymology and justification of the Persian counterpart, suggested mostly by the author. We hope that future editions of the dictionary will include counterparts in other languages and we would welcome any initiative in this direction.

While working on this dictionary, we gave utmost attention to the linguistic and terminological aspects of words, their morphological structure, and, in a broader scope, the mechanisms that govern a scientific language. In particular, affixes (prefixes and suffixes) constitute an essential element of technical terms in the Indo-European languages. They are therefore included with due consideration.

## 3. Chain of associated concepts

Apart from hyperlinks to the concepts occurring in definitions, the dictionary also guides the reader to families of associated concepts. For example, the term *diffraction* is hyperlinked to *diffusion*, *dispersion*, *distribution*, *scattering*, which cannot be interchanged. Similarly, *merge* leads to *fusion* and *coalesce*. Sometimes the trajectory conducts to notions that are not directly related to astronomy. For example, *wisdom* appears as an entry because of its semantic relation with *rational* (number), *rationalism*, *reason*, and *reasoning*. The introduction of these concepts in the dictionary is justified because they constitute the most basic building blocks of understanding and science. The dictionary gives careful attention to fundamentals for the sake of clarity and education. Another example is *secularism* which is associated with *secular* in the astronomical terms *secular acceleration*, *secular stability* and so forth. This aspect of the work is currently being expanded to all possible cases. This approach in the formation of a technical dictionary, though new, wishes to display the conceptual space in which the astronomical subset vocabulary operates.

## 4. Culture, flowers, dreams

Languages convey culture and diversity and should be preserved as such. Imagine a magnificent garden with lots of different flowers, each with its proper color, shape, and perfume; and compare it with another garden filled solely with one variety of flowers. No matter how lovely the latter may be, it looks desolate insofar as it dramatically lacks diversity. Similarly, imagine an Earth with one unique civilisation, where everyone would wear the same clothes, eat the same food, listen to the same music, read the same books and have the same dreams. It would be terribly dull and sterile.

In the present age of exponential scientific and technological developments, the languages which are incapable of expressing new concepts are unfortunately doomed to disappear. It would be a dramatic loss if historical languages, which have made important contributions to human culture and civilization, and therefore belong to a common human heritage, died out. In particular, Persian represents a culture whose roots date back to several millennia and that has impacted human culture in several ways. In fact the number of such rare living cultural species in the whole human civilization is scarce. As for French, it is one of the languages of this dictionary not only because the author is familiar with it, but also due to its prominent standing as a cultural and scientific language. Despite the fact that there is seemingly a lack of will and interest among French astrophysicists to coin concise French equivalents for modern concepts in their profession, it would be wise to create a reservoir of such terms. Lacking French equivalent terms is one thing, possessing such terms but not using them is another. As long as suitable French equivalents do not exist, people are obliged to use English terms in French. In the author's opinion, such a situation is not laudable. Creating French equivalents maintains the capability of the language to express the most modern concepts without undue borrowings from other languages and keeps its word forming mechanisms alive. Let us be clear. Languages borrow from each other and there is no problem with this provided that the number of loanwords does not exceed a certain threshold and that they do not affect the language structure. Otherwise this will lead to the paralysis of the receiver language, as has been experienced for example by Persian. After all, who can guarantee that English will remain the privileged language of science for ever? For several centuries the language of science and philosophy was Latin. Sometime later French had the upper hand in western cultures. And another language may take over in some decades. At any rate, a language can stay alive only by standing on its own feet.

## 5. Tie to Virtual Observatories

This project relates also to astronomical Virtual Observatories. These online databases mainly contain digital data archives for their multiple and exhaustive exploitation by users from public astronomical community. In addition, Virtual Observatories include dictionaries of nomenclatures, which are closely tied to astronomical concepts. Although Virtual Observatories recognize the necessity of efforts to define basic astronomical concepts and establish their reciprocal relations, so far they have mainly been confined to archiving observational data. The present dictionary could be an incipient contribution to cover and inter-relate the whole astronomical lexicon beyond subfields.

## 6. Epilogue

The astronomical part of this project benefited from my fellow astronomers' contributions, mainly at Paris Observatory [2]. The linguistics part and other associated aspects rely on many written sources [3] as well as precious advice from several individuals all over the world – Europe, Northern America, and Iran, as recognized in the *Acknowledgments* section of the dictionary. May this work be the seed of a wider endeavour to collect and create a fully international astronomical thesaurus including languages from all continents.

## 7. References

[1] Heydari-Malayeri, M., 2009: http://www.obspm.fr/dico
[2] http://aramis2.obspm.fr/~heydari/dictionary/?static=staticAcknowledgments&encoding=ISO-8859-15
[3] Some 300 references, see:
http://aramis2.obspm.fr/~heydari/dictionary/index.php?static=staticReferences&encoding=ISO-8859-15